\documentclass[twocolumn,showpacs,superscriptaddress,preprintnumbers,amsmath,amssymb,prc]{revtex4-2}

\usepackage{graphicx}
\usepackage{dcolumn}
\usepackage{bm}
\usepackage{float}
\usepackage{color}
\usepackage{CJK}
\usepackage{ulem}
\usepackage{array} 
\usepackage{makecell}
\usepackage{multirow}

\usepackage[colorlinks,linkcolor=blue,urlcolor=blue,citecolor=blue]{hyperref}

\usepackage{tikz,xcolor,hyperref}
\definecolor{lime}{HTML}{A6CE39}

\DeclareRobustCommand{\orcidicon}{
	\begin{tikzpicture}
	\draw[lime, fill=lime] (0,0) 
	circle [radius=0.16] 
	node[white] {{\fontfamily{qag}\selectfont \tiny ID}};
	\draw[white, fill=white] (-0.0625,0.095) 
	circle [radius=0.007];
	\end{tikzpicture}
	\hspace{-2mm}
}
\foreach \x in {A, ..., Z}{%
	\expandafter\xdef\csname orcid\x\endcsname{\noexpand\href{https://orcid.org/\csname orcidauthor\x\endcsname}{\noexpand\orcidicon}}
}

\begin{document}
\begin{CJK*} {UTF8} {gbsn}

\title{Temperature and density  effects on the two-nucleon momentum correlation function from  excited single nuclei}

\author{Ting-Ting Wang(王婷婷)}
\affiliation{Key Laboratory of Nuclear Physics and Ion-Beam Application (MOE), Institute of Modern Physics, Fudan University, Shanghai 200433, China}

\author{Yu-Gang Ma(马余刚)\footnote{Corresponding author: mayugang@fudan.edu.cn}}
\affiliation{Key Laboratory of Nuclear Physics and Ion-Beam Application (MOE), Institute of Modern Physics, Fudan University, Shanghai 200433, China}
\affiliation{Shanghai Research Center for Theoretical Nuclear Physics， NSFC and Fudan University, Shanghai 200438, China}

\author{De-Qing Fang(方德清)}
\affiliation{Key Laboratory of Nuclear Physics and Ion-Beam Application (MOE), Institute of Modern Physics, Fudan University, Shanghai 200433, China}
\affiliation{Shanghai Research Center for Theoretical Nuclear Physics， NSFC and Fudan University, Shanghai 200438, China}

\author{Huan-Ling Liu(刘焕玲)}
\affiliation{Shanghai Institute of Applied Physics, Chinese Academy of Sciences, Shanghai 201800, China}

\date{\today}

\begin{abstract}
Two-nucleon momentum correlation functions are investigated for different single thermal sources  at given initial temperature $(T)$ and density $(\rho)$. 
To this end, the space-time evolutions of various  single excited nuclei at $T$ $= 1 - 20$ $MeV$ and $\rho$ = 0.2 - 1.2 $\rho_0$ are simulated by using the thermal isospin-dependent quantum molecular dynamics $(ThIQMD)$ model. Momentum correlation functions of identical proton-pairs ($C_{pp}(q)$) or neutron-pairs ($C_{nn}(q)$)  at small relative momenta are calculated by  $Lednick\acute{y}$ and $Lyuboshitz$ analytical method.
The results illustrate that $C_{pp}(q)$ and $C_{nn}(q)$ 
are sensitive to the source size ($A$) at lower $T$ or higher $\rho$, but almost not at higher $T$ or lower $\rho$. And the sensitivities become stronger for smaller source. 
Moreover, the $T$, $\rho$ and $A$ dependencies of the Gaussian source radii are also extracted by fitting the two-proton  momentum correlation functions, and 
the results are consistent with the above conclusions. 
\end{abstract}


\maketitle

\section{Introduction}
Properties of nuclear matter is one of the most interesting topics in heavy-ion physics ~\cite{GGiuliani2014,BALi2008,BBorderie2008,CWMa2018} and lots of works have been done around zero temperature, including the nuclear equation of state $(EOS)$.  
However, the studies on properties of nuclear matter at finite temperatures are relatively limited. Many previous works mainly focus on the temperature dependence of hot nuclear matter and the nuclear liquid-gas phase transition $(LGPT)$~\cite{Liuhuanling,JEFinn1982,PJSiemens1983,YGMa1997,YGMa1999,PChomaz2000,JBNatowitz2002,YGMa2005,BBorderie2019,zhangzhenfang2018}, the ratio between shear viscosity over entropy density $\left(\eta/s\right)$~\cite{NAuerbach2009,NDDang2011,DQFang2014,XGDeng2016,DMondal2017}, 
as well as the nuclear giant dipole resonance~\cite{ABracco1989,PFBortignon1991,OWieland2006} etc. Among above works, the relationship between the phase transition 
temperature and the source size has been investigated~\cite{Liuhuanling}. In Ref.~\cite{Liuhuanling}, the finite-size scaling effects on nuclear liquid-gas phase transition 
probes are investigated by studying de-excitation processes of the thermal sources by the isospin-dependent quantum molecular dynamics model (IQMD).  
Several probes, including the total multiplicity derivative, 
second moment parameter, 
intermediate mass fragment multiplicity, 
$Fisher^{,}s$ power-law exponent 
as well as nuclear $Zipf^{,}s$ law exponent of Ma \cite{YGMa1999} were explored, and  the phase transition temperatures were then obtained. 
Recently, the deep neural network has also been used to determine the nuclear liquid gas phase transition \cite{WangRui} and  to estimate the temperature of excited nuclei by 
the  charge multiplicity distribution of emitted fragments~\cite{YDSong2021}. The latter work proposed that  the charge multiplicity distribution can be used as a thermometer of heavy-ion collisions.

Considering that the intermediate-state at high temperature and density in the evolution process of nuclear reactions cannot be directly measured, one always explore properties of nuclear matter and the dynamical description of heavy-ion collisions through the analysis of the final-state products. 
As well known, the two-particle momentum correlation function in the final-state has been  extensively used  as a probe of the space-time properties and characteristics of the emission source ~\cite{RKotte2005,RGhetti2003,DGourio2000}. 
The two-proton momentum correlation function  has been explored systematically by a lot of experiments  as well as different models, 
several reviews can be found in Refs.~\cite{Boal,Heinz,Lisa,Verde}. In various studies on the momentum correlation function, impacts of the impact parameter, the total momentum of nucleon pairs, the isospin of the emission source, the nuclear symmetry energy, the nuclear equation of state $(EOS)$ as well as the in-medium nucleon-nucleon cross section 
have been discussed in literature~\cite{WGGong1991,Verde2,YGMa2006,RGhetti2004,LWChen2003,wtt2018,wtt2019}. Even more, nuclear structure effects 
were also carefully investigated, such as the effects from binding energy and separation energy of the nucleus \cite{WeiYB}, density distribution of 
valence neutrons in neutron-rich nuclei \cite{CaoXG},  as well as high momentum tail of the nucleon-momentum distribution \cite{WeiGF} etc.   
Two-proton momentum correlation function was also constructed in few-body reactions  as well as $\alpha$-clustered nucleus induced collisions \cite{Fang,Huang1,Huang2,Shen,HeJJ}. 
In addition, momentum correlation function between two light charged particles also offers a unique tool to investigate dynamical expansion of the reaction zone~\cite{wtt2019}.

Here we extend the momentum correlation method of final-state interaction to study the time-spatial information of the finite-temperature nuclear systems which have different initial density. 
The purpose of the present paper is to systematically investigate  the relationship between  two-particle momentum correlation functions and system parameters, such as the source-temperature, density as well as system-size in a framework of the thermal isospin-dependent quantum molecular dynamics $(ThIQMD)$ model ~\cite{DQFang2014,Liuhuanling,zhangzhenfang2018}. In addition, the Gaussian source radii are quantitatively extracted by  assumption of Gaussian source fits to the momentum correlation function distributions.
In this article, the evolution process of excited nuclear sources at given initial temperatures varying 
from $1$ $MeV$ to $20$ $MeV$ are studied. 
The present work selects six different nuclear systems with similar ratio of neutron to proton numbers, $i.e.$, $N/Z \sim 1.3$, which include $(A, Z)$ = $(36, 15)$, $(52, 24)$, $(80, 33)$, $(100, 45)$, $(112, 50)$, and $(129, 54)$ nuclei. Then,  Lednick$\acute{y}$-Lyuboshitz theoretical approach~\cite{lednicky2006} is applied for calculating two-particle momentum correlation functions which are constructed based on phase-space information from the evolution process of single excited nuclear sources by the $ThIQMD$ model.

The rest of this article is organized as follows. In Section $II$, we firstly describe the thermal isospin-dependent quantum molecular dynamics model ~\cite{DQFang2014,zhangzhenfang2018}, 
then briefly introduce the momentum correlation technique using $Lednick\acute{y}$ and $Lyuboshitz$ analytical formalism. 
In Section $III$, we show the results of the $ThIQMD$ plus the $LL$ method for the source-temperature dependence of 
two-particle momentum correlation function.  The two-particle momentum correlation functions of different  system-sizes at different initial densities are systematically discussed. The detailed analysis of the extracted Gaussian source radii are presented under  different source-temperature and density. Furthermore, the momentum correlation function of two-neutron is also analyzed. Finally, Section $IV$ gives a summary of the paper.

\section{MODELS AND FORMALISM}
\subsection{THE ThIQMD MODEL}
In this paper, the thermal isospin-dependent Quantum Molecular Dynamics  transport model is used as the event generator, which has been applied successfully to study the $LGPT$~\cite{Liuhuanling,YDSong2021}.
In the following discussion, we introduce this model briefly. As well known, isospin-dependent Quantum Molecular Dynamics ($IQMD$) model was used to describe the collision process between two nuclei. 
The Quantum Molecular Dynamics transport model is a $n$-body transport theory, which describes heavy-ion reaction dynamics from intermediate to relativistic energies~\cite{Aichelin1987,Aichelin1991,Peilert1989,Feng2018}.   
In the present work, we use a single excited source in the $ThIQMD$ which is different from the traditional $IQMD$. Usually, the ground state of the initial nucleus is considered to be $T = 0$ $MeV$ 
in the traditional $IQMD$ model. However, the $ThIQMD$ model developed by Fang, Ma, and Zhou  in Ref.~\cite{DQFang2014} 
is used to simulate single thermal  source at different temperatures and densities. 

The main  parts of $QMD$ transport model include the following issues: the initialization of the projectile and the target, nucleon propagation under the effective potential, 
the collisions between the nucleons in the nuclear medium and the Pauli blocking effect. In the $ThIQMD$, instead of using the Fermi-Dirac distribution for $T = 0$ $MeV$ with the nucleon's maximum momentum limited by 
$P_{F}^{i}(\vec{r}) = \hbar\left[3 \pi^{2} \rho_{i}(\vec{r})\right]^{1 / 3}$, the initial momentum of nucleons is sampled by the Fermi-Dirac distribution at finite temperature:
\begin{equation}
n\left(e_{k}\right) = \frac{g\left(e_{k}\right)}{e^{\frac{e_{k}-\mu_i}{T}}+1}, 
\end{equation}
where the kinetic energy $e_{k} = \frac{p^{2}}{2 m}$, $p$ and $m$ is the momentum and mass of the nucleon,
respectively. $g\left(e_{k}\right) = \frac{V}{2 \pi^{2}}\left(\frac{2 m}{\hbar^{2}}\right)^{\frac{3}{2}} \sqrt{e_{k}}$ represents the state density with the volume of the source
$V = \frac{4}{3} \pi r^{3}$ where $r = r_{V} A^{\frac{1}{3}}$ ($r_V$ is a parameter to adjust the initial density). 

In addition, the chemical potential $\mu_i$ is determined by the following equation:
\begin{equation}
\frac{1}{2 \pi^{2}}\left(\frac{2 m}{\hbar^{2}}\right)^{\frac{3}{2}} \int_{0}^{\infty} \frac{\sqrt{e_{k}}}{e^{\frac{e_{k}-\mu_i}{T}}+1} d e_{k} = \rho_{i}.
\end{equation}
where $i$ = $n$ or $p$ refer to the neutron or proton.

In the $ThIQMD$ model, the interaction potential is also represented by the form as follows:
\begin{equation}
U = U_{Sky} + U_{Coul} + U_{Yuk} + U_{Sym} + U_{MDI}, 
\end{equation}
where $U_{Sky}$, $U_{Coul}$, $U_{Yuk}$, $U_{Sym}$, and $U_{MDI}$ are the density-dependent Skyrme potential, the Coulomb potential, 
the surface Yukawa potential, the isospin asymmetry potential, and the momentum-dependent interaction, respectively.
Among these potentials,  the Skyrme potential, the Coulomb potential and the momentum-dependent interaction can be written as follows:
\begin{equation}
U_{Sky}=\alpha (\frac \rho {\rho _0})+\beta (\frac \rho {\rho
_0})^\gamma, 
\end{equation}
where $\rho$ and $\rho _{_0}$ are total nucleon density and its normal value at the ground state, $i.e.$, $0.16$ ${fm}^{-3}$, respectively. The above parameters $\alpha$, $\beta$, and $\gamma$ with an incompressibility parameter $K$ are related to the nuclear equation of state~\cite{BALi,Zhang-Li17,Cai17,Zhang17,Wei,Jiang,Xu}.
\begin{equation}
U_{Sym} = C_{sym} \frac{\left(\rho_{n}-\rho_{p}\right)}{\rho_{0}}\tau_{z},
\end{equation} 
\begin{equation}
U_{Coul} = \frac{1}{2}\left(1-\tau_{z}\right)V_{c},
\end{equation} 
where $\rho_{n}$ and $\rho_{p}$ are neutron and proton densities, respectively, $\tau_{z}$ is the $z$-th component of the isospin degree of freedom for the nucleon, which equals $1$ or $-1$ for a neutron or proton, respectively, and $C_{sym}$ is the symmetry energy coefficient. 
$U_{Coul}$ is the Coulomb potential where $V_{c}$ is its parameter for protons. 
\begin{equation}
U_{MDI} = \delta \cdot \ln ^{2}\left(\epsilon \cdot(\Delta p)^{2}+1\right) \cdot \frac{\rho}{\rho_{0}}, 
\end{equation}
where $\Delta p$ is the relative momentum, $\delta$ and $\epsilon$ can be found in Refs.~\cite{Aichelin1987,Aichelin1991}. Their values of the above potential parameters are all listed in Table $I$:

\begin{table}[!htbp]
\caption
{The value of the interaction potential parameters.}
\begin{tabular}{cccccc}
\hline
\hline
$\alpha$      &      $\beta$      &      $\gamma$      &      $K$            &      $\delta$      &      $\epsilon$\\ \hline   
$(MeV)$         &     $(MeV)$         &        &     $(MeV)$        &     $(MeV)$         &     ($(GeV/ {c})^{-2}$)\\ \hline 
$-390.1$        &     $320.3$         &  $1.14$          &      $200$          &  $1.57$          &      $500$\\ \hline
\hline
\end{tabular}
\end{table}

\subsection{LEDNICK$\acute{Y}$ AND LYUBOSHITZ ANALYTICAL FORMALISM}

Next, we briefly review the method  for the two-particle momentum correlation function proposed by Lednick$\acute{y}$ and Lyuboshitz~\cite{lednicky2006,lednicky2009,lednicky2008}.
The momentum correlation technique in nuclear collisions is based on the principle as follows: when they are emitted at small relative momentum, the two-particle momentum correlation is determined by the space-time characteristics of the production processes owing to the effects of quantum statistics $(QS)$ and final-state interactions $(FSI)$~\cite{Koonin1977,Lednicky2007}. Therefore, the two-particle momentum correlation function can be expressed through a square of the symmetrizied Bethe-Salpeter amplitude averaging over the four coordinates of the emitted particles and the total spin of the two-particle system, which represents the continuous spectrum of the two-particle state. 

In this theoretical approach, the final-state interactions of the particle pairs is assumed independent in the production process. According to the conditions in Ref.~\cite{Lednicky1996}, the correlation function of two particles can be written as the expression:
\begin{equation}
\textbf{C}\left(\textbf{k}^*\right) = \frac{\int
\textbf{S}\left(\textbf{r}^*,\textbf{k}^*\right)
\left|\Psi_{\textbf{k}^*}\left(\textbf{r}^*\right)\right|^{2}d^{4}\textbf{r}^*}
{\int
\textbf{S}\left(\textbf{r}^*,\textbf{k}^*\right)d^{4}\textbf{r}^*},
\end{equation}
where $\textbf{r}^* = \textbf{x}_{1}-\textbf{x}_{2}$ is the relative distance of the two particles  in the pair rest frame $\left(PRF\right)$ at their kinetic freeze-out, $\textbf{k}^*$ is half of the relative momentum between two particles in the $PRF$, $\textbf{S}\left(\textbf{r}^*,\textbf{k}^*\right)$ is the probability to emit a particle pair with given $\textbf{r}^*$ and $\textbf{k}^*$, $i.e.$, the source emission function, and $\Psi_{\textbf{k}^*}\left(\textbf{r}^*\right)$ is the equal-time $\left(t^* = 0\right)$ reduced Bethe-Salpeter amplitude which can be approximated by the outer solution of the scattering problem in the $PRF$~\cite{lednicky1982,Star-nature}. This approximation is valid on condition $\left|t^{*} \right|\ll m\left(r^{*}\right )^{2}$, which is well fulfilled for sufficiently heavy particles like protons or kaons and reasonably fulfilled even for pions~\cite{lednicky2009}.
In the above limit, the asymptotic solution of the wave function of the two charged particles approximately takes the expression:
\begin{multline}
\Psi_{\textbf{k}^*}\left(\textbf{r}^*\right) = e^{i\delta_{c}}\sqrt{A_{c}\left(\lambda \right)} \times\\
\left[e^{-i\textbf{k}^*\textbf{r}^*}F\left(-i\lambda,1,i\xi\right)+f_c\left(k^*\right)\frac{\tilde{G}\left(\rho,\lambda \right)}{r^*}\right].
\end{multline}
In the above equation, $\delta_{c} =$ $arg$ $\Gamma\left(1+i\lambda\right)$ is the Coulomb $s$-wave phase shift with $\lambda = \left(k^*a_c\right)^{-1}$ where $a_{c}$ is the two-particle Bohr radius, $A_c\left(\lambda \right) = 2\pi\lambda \left[\exp\left(2\pi\lambda \right)-1\right]^{-1}$ is the Coulomb penetration factor, and its positive (negative) value corresponds to the repulsion (attraction). $\tilde{G}\left(\rho,\lambda \right) = \sqrt{A_{c}\left(\lambda \right)}\left[G_0\left(\rho,\lambda \right)+iF_0\left(\rho,\lambda \right)\right]$ is a combination of regular $\left(F_0\right)$ and singular $\left(G_0\right)$ $s$-wave Coulomb functions~\cite{lednicky2009,lednicky2008}. $F\left(-i\lambda,1,i\xi\right) = 1+\left(-i\lambda\right)\left(i\xi\right)/1!^{2}+\left(-i\lambda\right)\left(-i\lambda+1\right)\left(i\xi\right)^{2}/2!^{2}+\cdots$ is the confluent hypergeometric function with $\xi = \textbf{k}^*\textbf{r}^*+\rho$, $\rho = k^*r^*$.
\begin{equation}
f_c\left(k^*\right) = \left[ K_{c}\left(k^*\right)-\frac{2}{a_c}h\left(\lambda \right)-ik^*A_{c}\left(\lambda \right)\right]^{-1}
\end{equation}
is the $s$-wave scattering amplitude renormalizied by the long-range Coulomb interaction, with $h\left(\lambda \right) = \lambda^{2}\sum_{n=1}^{\infty}\left[n\left(n^2+\lambda^2\right)\right]^{-1}-C-\ln\left[\lambda \right]$ where $C$ = 0.5772 is the Euler constant.
$K_{c}\left(k^*\right) = \frac{1}{f_0} + \frac{1}{2}d_0k^{*^2} + Pk^{*^4} + \cdots$ is the effective range function, where $d_{0}$ is the effective radius of the strong interaction, $f_{0}$ is the scattering length and $P$ is the shape parameter. The parameters of the effective range function are important parameters characterizing the essential properties of the $FSI$, and can be extracted from the correlation function measured experimentally~\cite{Erazmus1994,Arvieux1974,Star-nature,wtt2019}.

For $n$-$n$ momentum correlation functions which include uncharged particle, only the short-range particle interaction works. For $p$-$p$ momentum correlation functions, both the Coulomb interaction and the short-range particle interaction dominated by the $s$-wave interaction are taken into account. 

\section{ANALYSIS AND DISCUSSION}

Within the framework of the thermal isospin-dependent quantum molecular dynamics  model~\cite{DQFang2014,Liuhuanling,zhangzhenfang2018}, 
the two-particle momentum correlation functions are calculated by using the phase-space information from the freeze-out stage of the excited nuclear source 
at an initial temperature varying from $1$ $MeV$ to $20$ $MeV$ and/or density varying from $\rho = 0.2 \rho_{0}$ to $1.2 \rho_{0}$. This work performs calculations for thermal source systems with different mass including $(A, Z)$ $=$ $(36, 15), (52, 24), (80, 33), (100, 45), (112, 50),$ and $(129, 54)$. 
\begin{figure}[!htbp]
 \includegraphics[width=\linewidth]{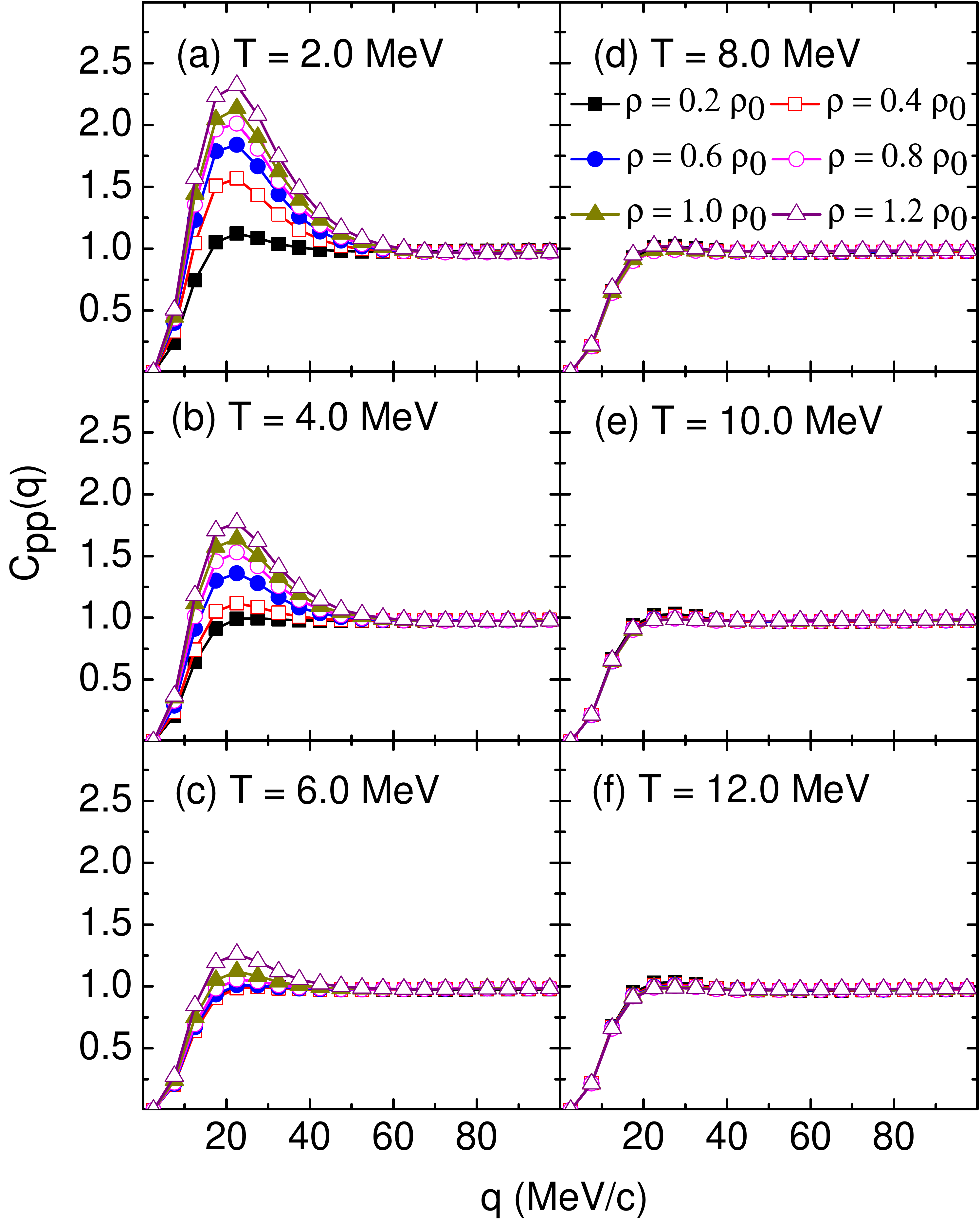}
 \centering
 \caption{
The proton-proton momentum correlation function  $(C_{pp}(q))$ at different densities ($i.e.$, $0.2 \rho_{0}$, $0.4 \rho_{0}$, $0.6 \rho_{0}$, $0.8 \rho_{0}$, $1.0 \rho_{0}$, and $1.2 \rho_{0}$) for the smaller nucleus ($A$=36, $Z$=15) with fixed source-temperatures $T = 2$ $MeV$ (a), $4$ $MeV$ (b), $6$ $MeV$ (c), $8$ $MeV$ (d), $10$ $MeV$ (e) and $12$ $MeV$ (f), respectively. The freeze-out time is taken to be  $200$ $fm/c$.  
}
 \label{fig1}
\end{figure}
\begin{figure}[!htbp]
 \includegraphics[width=\linewidth]{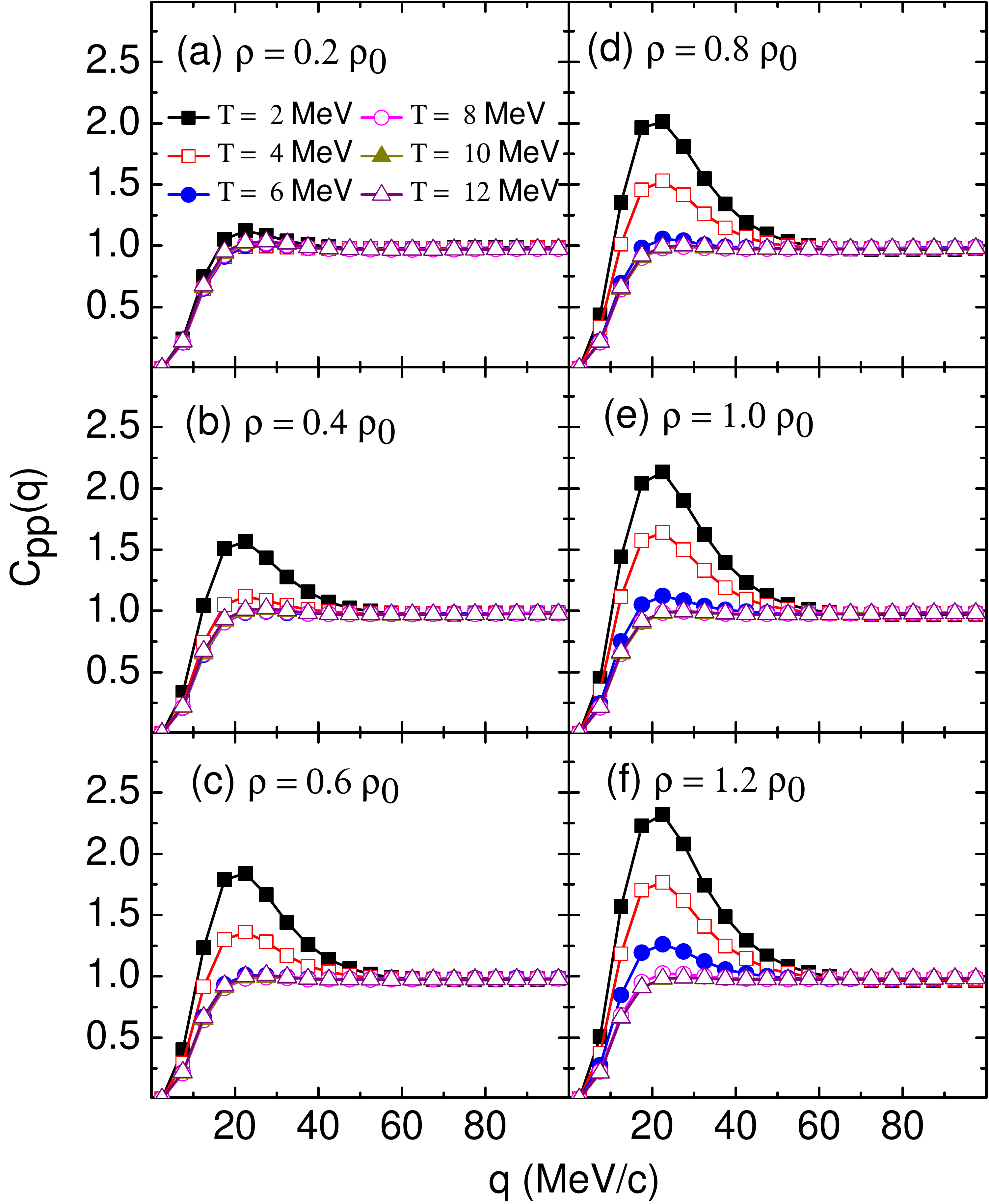}
 \centering
 \caption{
Similar to Fig.~\ref{fig1}, but  at different source-temperatures ($T$ = 2, 4, 6, 8, 10 and 12 $MeV$) with different fixed densities, namely $\rho$ = $0.2 \rho_{0}$ (a), 0.4 $\rho_{0}$ (b), 0.6 $\rho_{0}$ (c), 0.8 $\rho_{0}$ (d), 1.0 $\rho_{0}$ (e), and $1.2 \rho_{0}$ (f). 
 }
 \label{fig2}
\end{figure}
\begin{figure}[!htbp]
 \includegraphics[width=\linewidth]{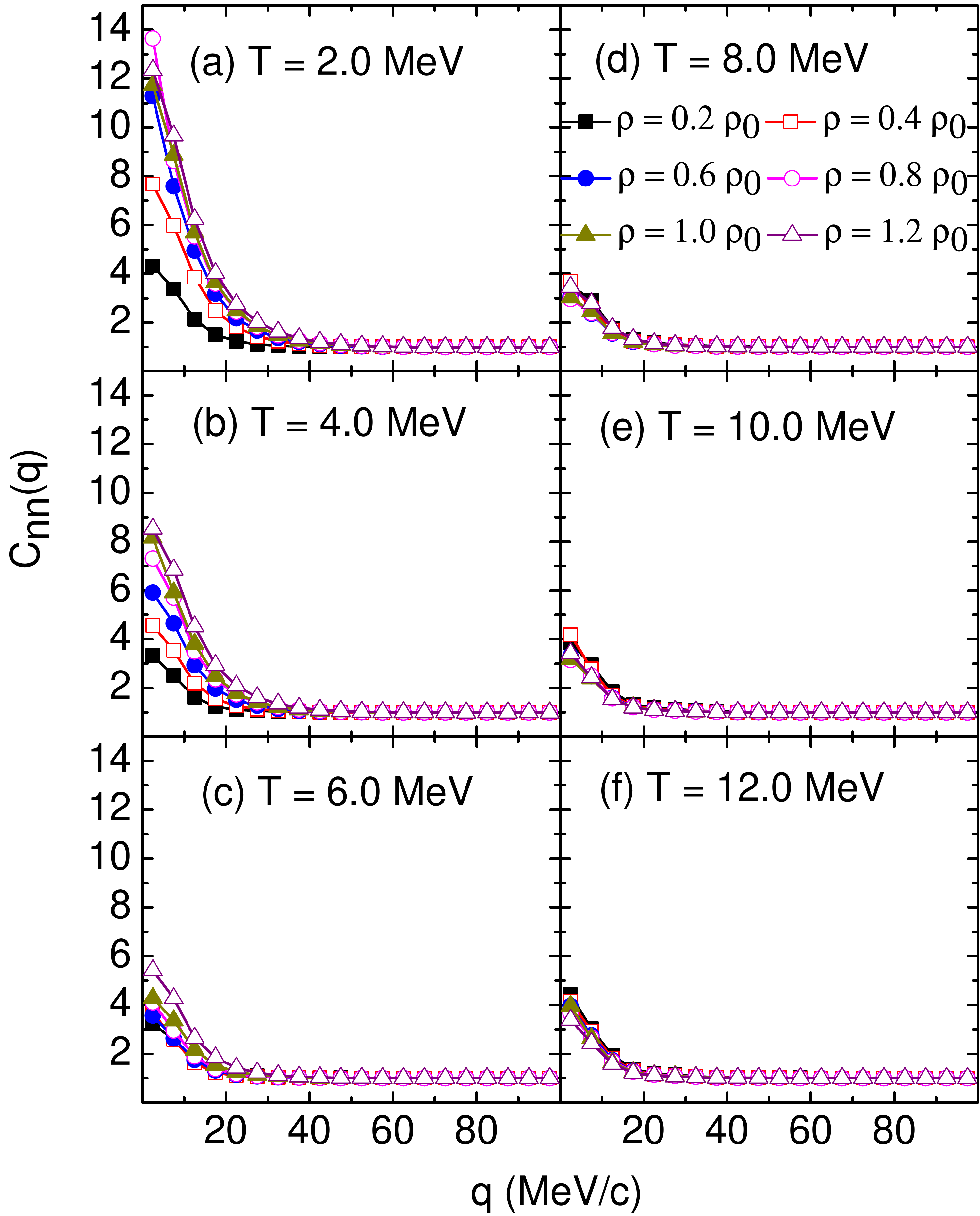}
 \centering
 \caption{
The neutron-neutron ($n$-$n$) momentum correlation functions $(C_{nn}(q))$ in the same conditions as Fig.~\ref{fig1}.
 }
 \label{fig3}
\end{figure}

\begin{figure}[!htbp]
 \includegraphics[width=\linewidth]{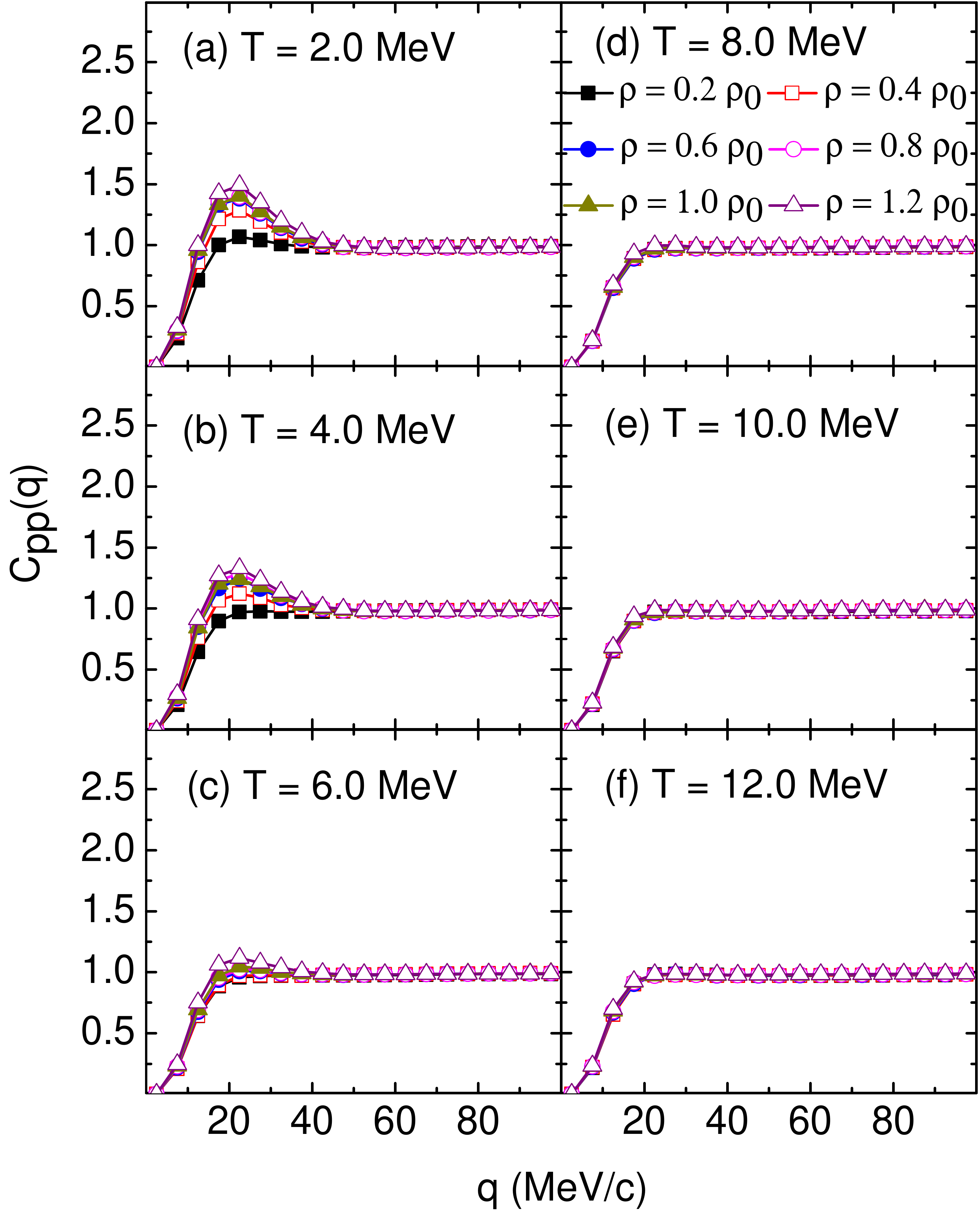}
 \centering
 \caption{
 Same to Fig.~\ref{fig1}, but for a larger system ($A$=129, $Z$=54).
 }
 \label{fig4}
\end{figure}

\begin{figure*}[!htbp]
 \includegraphics[width=\linewidth]{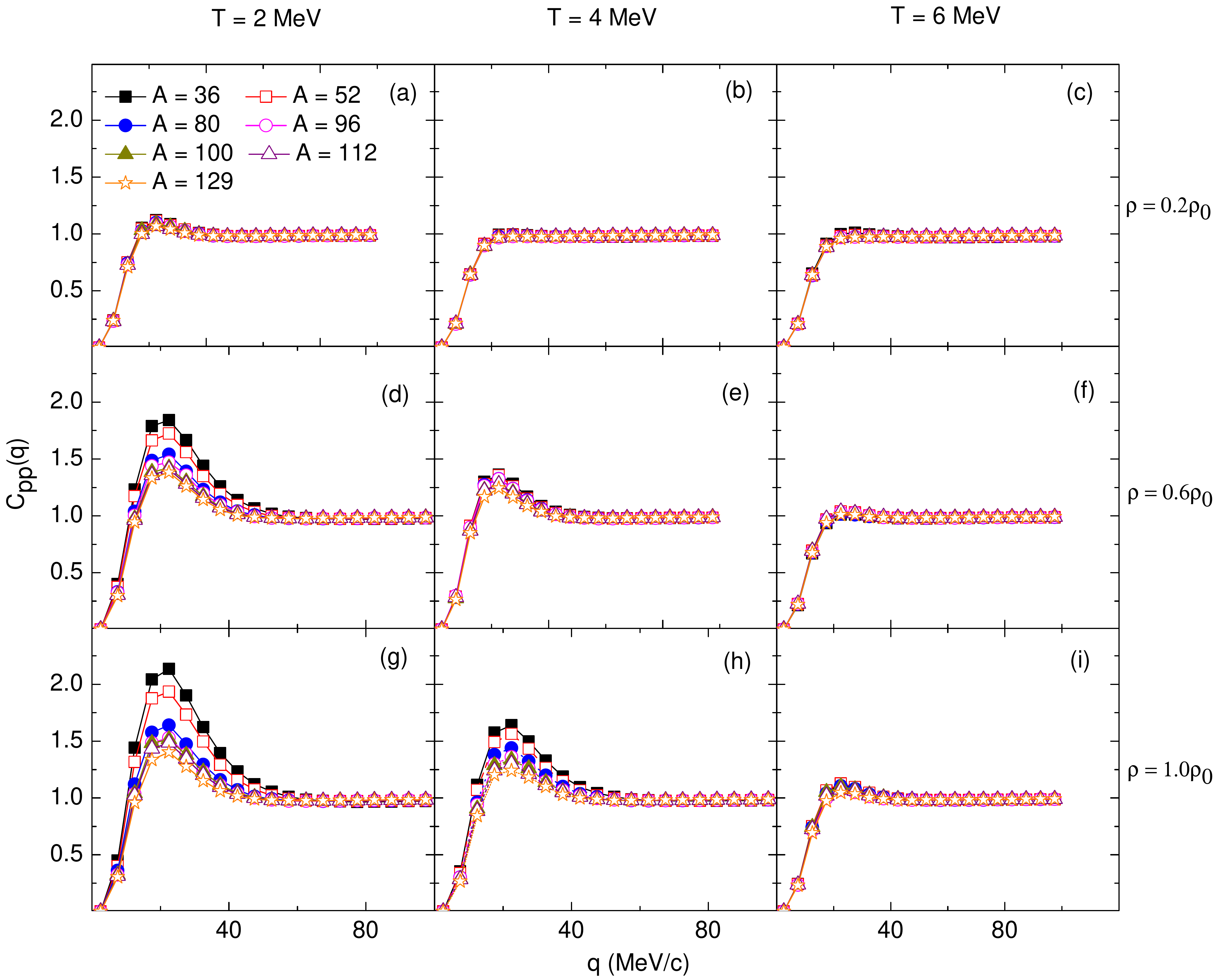}
 \centering
 \caption{
$C_{pp}(q)$ of different source size systems at fixed temperatures (i.e., from left column to right column, they correspond to $T$ = 2, 4 and 6 $MeV$, respectively) or fixed densities (i.e., from top row to bottom row, they correspond to  $\rho =$ $0.2 \rho_{0}$, $0.6 \rho_{0}$, $1.0 \rho_{0}$, respectively). 
 }
 \label{fig5}
\end{figure*}

\begin{figure}[!htbp]
 \includegraphics[width=\linewidth]{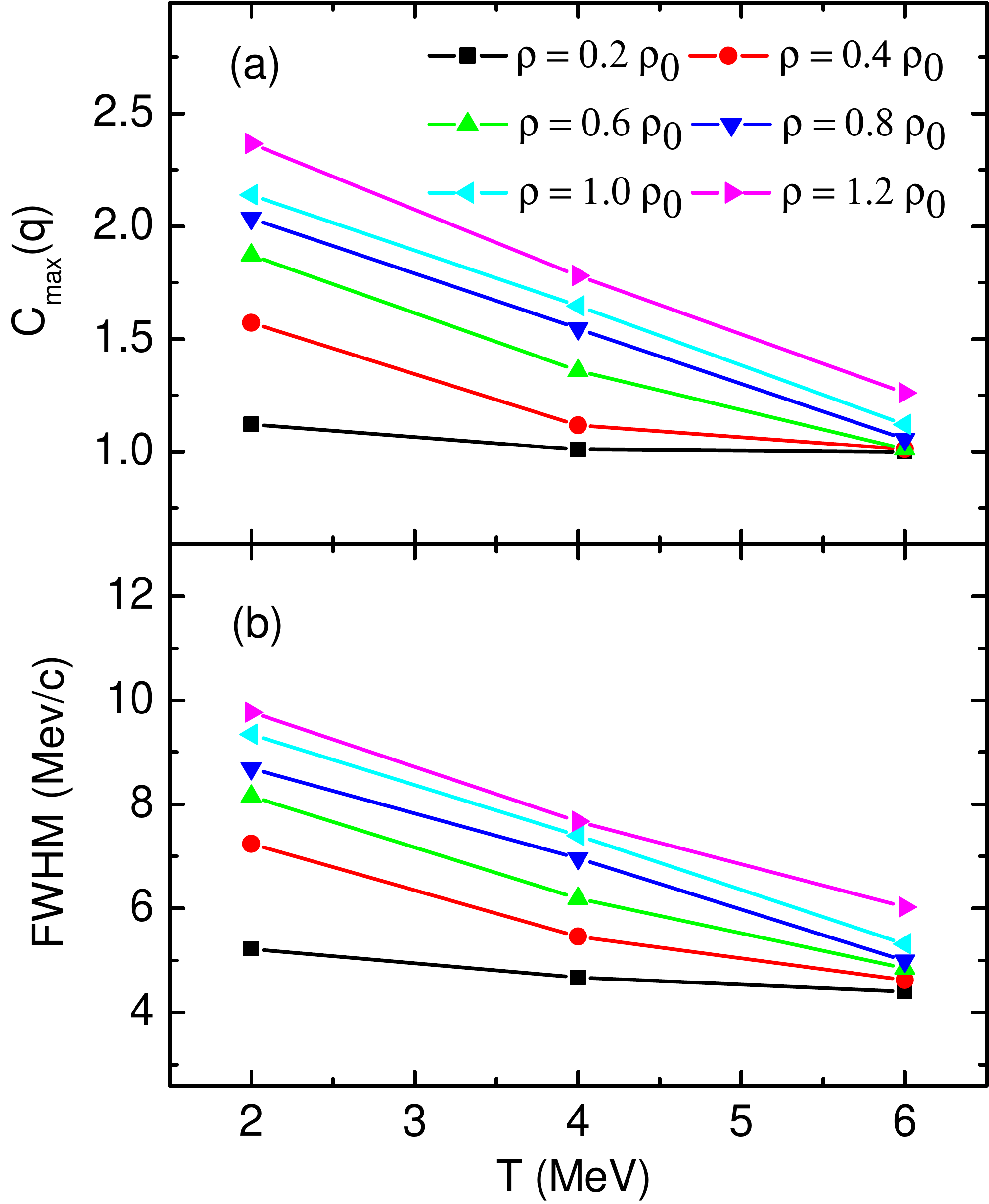}
 \centering
 \caption{
Source-temperature $T$ 
dependencies of $C_{max}(q)$ (a) and of $FWHM$ (b) of $C_{pp}(q)$  distributions at different densities ($0.2\rho_{0}$ - $1.2\rho_{0}$) for the ($A$ = 35, $Z$ = 16) system.}
 \label{fig6}
\end{figure}

\begin{figure}[!htbp]
 \includegraphics[width=0.9\linewidth]{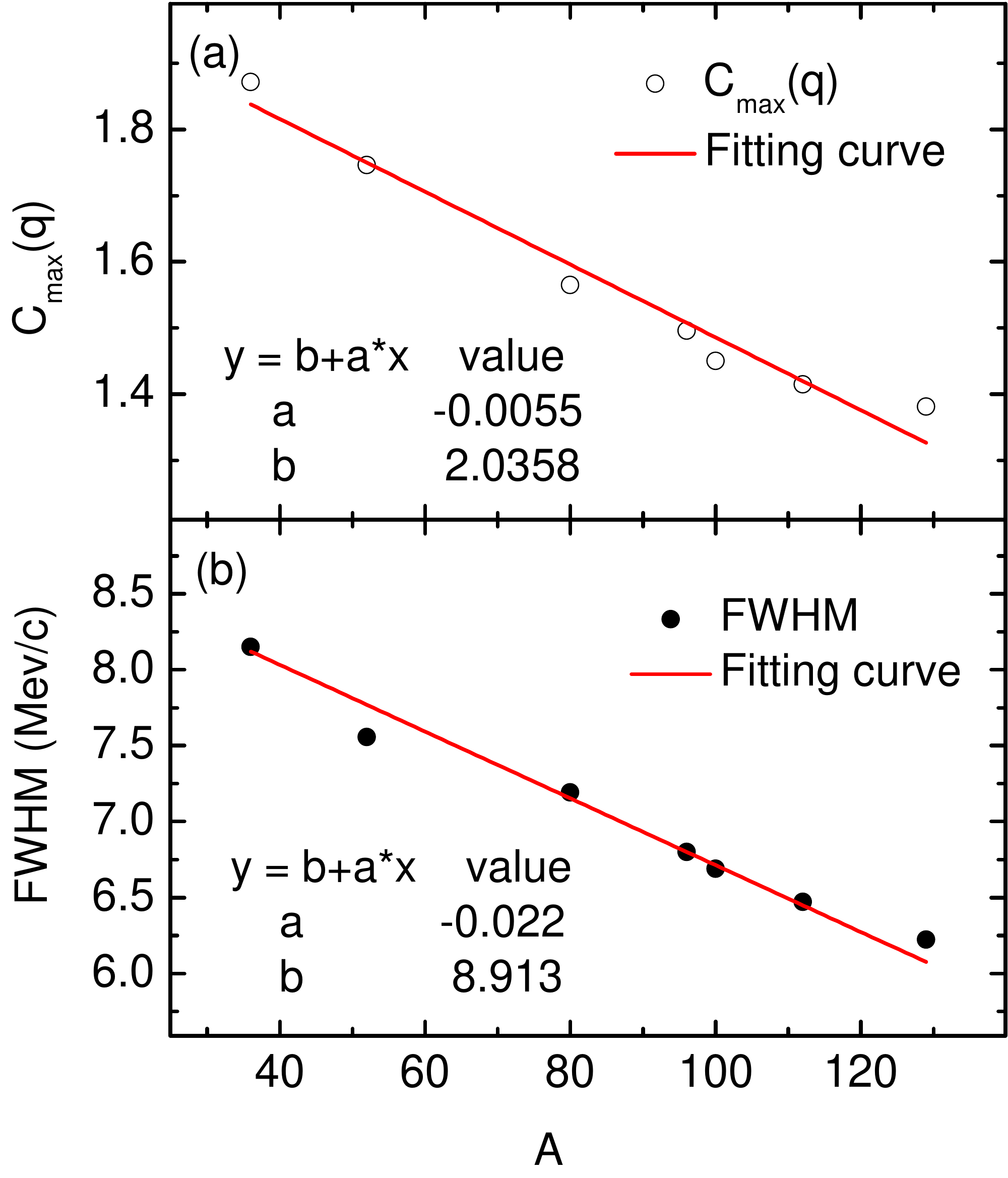}
 \centering
 \caption{
$C_{max}(q)$ (a) and $FWHM$ (b) for different source-size systems at given $T$ = 2 $MeV$ and $\rho = 0.6\rho_{0}$.}
 \label{fig7}
\end{figure}

\begin{figure}[!htbp]
 \includegraphics[width=0.9\linewidth]{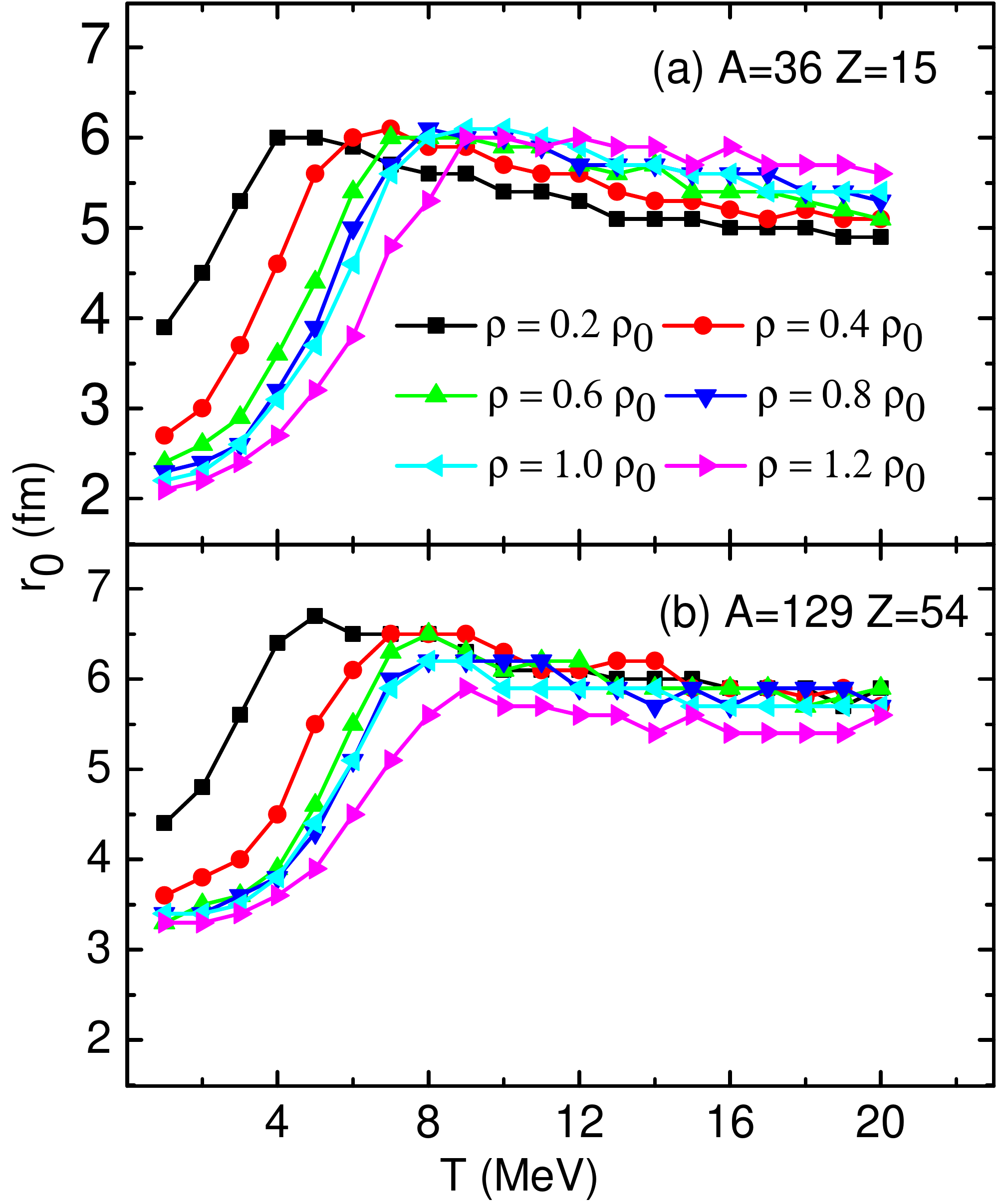}
 \centering
 \caption{
   Gaussian source radius as a function of temperature 
   at different densities ($\rho =$ $0.2\rho_{0}$, $0.4 \rho_{0}$, $0.6 \rho_{0}$, $0.8 \rho_{0}$, $1.0 \rho_{0}$, $1.2 \rho_{0}$) for a fixed source size. 
Panel (a) and (b) correspond to the smaller source size with ($A$ = 36, $Z$ = 15)  and the larger source size with ($A$ = 129, $Z$ = 54), respectively.
 }
 \label{fig8}
\end{figure}

\begin{figure}[!htbp]
 \includegraphics[width=\linewidth]{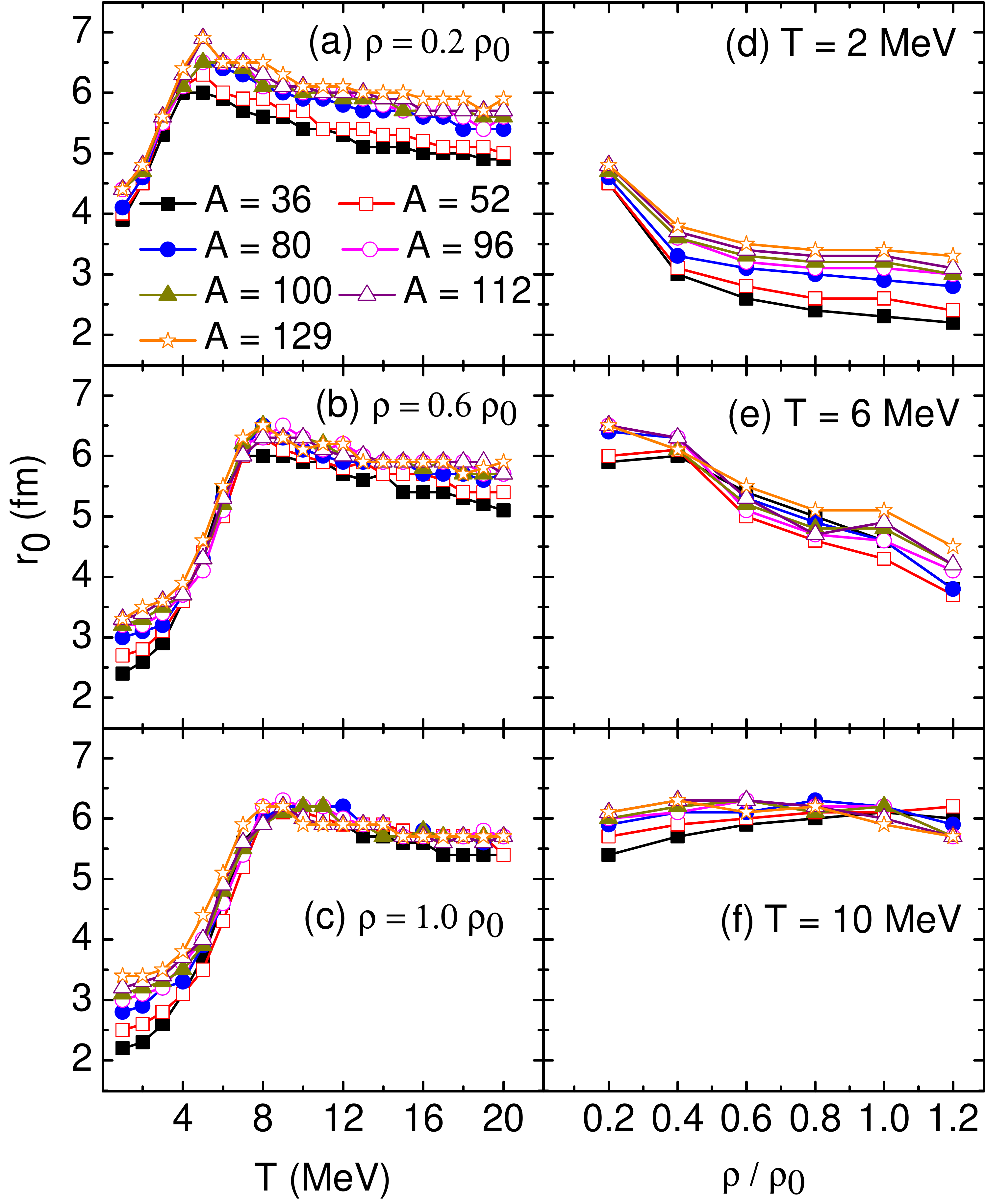}
 \centering
 \caption{
Gaussian source radius as a function of temperature or density at different source-size systems. Left and right columns correspond to $r_0$ at different densities, i.e $\rho =$  0.2 $\rho_0$ (a), 0.6 $\rho_0$  (b), and 1.0 $\rho_0$ (c) as well as different temperatures, i.e. $T =$ $2$ (d), $6$ (e), and $10$ (f) $MeV$,  respectively.
 }
 \label{fig9}
\end{figure}

We firstly calculated the proton-proton momentum correlation function $C_{pp}(q)$ for finite-size systems at temperatures ranging from $1$ to $20$ $MeV$. 
In Fig.~\ref{fig1}, the results of $C_{pp}(q)$ for temperature of $2$, $4$, $6$, $8$, $10$ and $12$ $MeV$ at different values of density ($0.2\rho_{0}$ - $1.2\rho_{0}$) are presented. 
The proton-proton momentum correlation function exhibits a peak at relative momentum $q$ $=$ $20$ $MeV/c$, 
which is due to the strong final-state $s$-wave attraction together with the suppression at lower relative momentum as a result of Coulomb repulsion and 
the antisymmetrization wave function between two protons. The shape of the two-proton momentum correlation functions is consistent with many previous experimental data 
in heavy-ion collisions, eg. Ref.~\cite{Ghetti2000}. For protons which are emitted from the lower temperature ($T$ $<$ $8$ $MeV$) source in Fig.~\ref{fig1} (a)-(c), 
the general trend is very similar. The figure shows that $C_{pp}(q)$ increases as $\rho$ increases for fixed $T$ ($T$ $<$ $8$ $MeV$). 
The increase of the density indicates that the geometrical size becomes smaller for a source with fixed neutrons and protons, which makes the strength of the momentum correlation function stronger. Finally, the $p$-$p$ momentum correlation function becomes almost one at $q$ $>$ $60$ $MeV/c$. 
For larger $T$ ($T$ $>$ $8$ $MeV$) in Fig.~\ref{fig1} (d)-(f), the difference of $C_{pp}(q)$ between different densities  becomes smaller. 
From Fig.~\ref{fig1}, it is found that the $C_{pp}(q)$ almost keep the same above $T = 8$ $MeV$ for different densities and the $p$-$p$ momentum correlation function becomes almost unique above  approximately $q = 30$ $MeV/c$. 
It indicates that the emitted proton is not affected by the change of density when the source temperature beyond certain value ($T \approx 8$ $MeV$ in present work). 
In order to understand which one of the two factors ($i.e.$, temperature and density) has larger influence, the two-particle momentum correlation 
in fig.~\ref{fig2} is plotted by exchanging of the two input parameters. From fig.~\ref{fig2}, we can intuitively observe dependence of the two-particle momentum correlation on the source temperature.
The dependence of $C_{pp}(q)$ on the source temperature is stronger than on density. In other words, the $C_{pp}(q)$ is more sensitive to $T$ than to density $\rho$. 
In addition, for larger $\rho$ from fig.~\ref{fig2} (a) to (f), the difference of $C_{pp}(q)$ between different densities becomes bigger.
Next, we explore whether the phenomenon exists in momentum correlation functions for the uncharged-particle pairs. Fig.~\ref{fig3} presents the neutron-neutron momentum 
correlation functions ($C_{nn}(q)$) for temperature of $2$, $4$, $6$, $8$, $10$ and $12$ $MeV$ at different values of density, respectively. For neutron-neutron momentum 
correlation function, it peaks at $q \approx 0$ $MeV/c$ caused by the $s$-wave attraction. Although the $C_{nn}(q)$ has different shape compared with the $p$-$p$ momentum 
correlation function, it has the similar dependence on the source temperature and density. The similar trend in $C_{pp}(q)$ and $C_{nn}(q)$ shows the close emission mechanism 
in the evolution process. 

Fig.~\ref{fig4} shows the results of a larger system at different source-temperature and density, and a similar behavior of $C_{pp}(q)$ is demonstrated. 
We also observe that the proton-proton momentum correlation in larger-size system ($(A, Z)=(129, 54)$) in Fig.~\ref{fig4} 
becomes weaker in comparison with the smaller-size source ($(A, Z)=(36, 15)$) in Fig.~\ref{fig1}. 
In view of the above phenomenon, Fig.~\ref{fig5} describes the relationship between system-size and momentum correlation function in more details. 
The decreasing of $C_{pp}(q)$ as the system-size increasing for a fixed value of $T$ or $\rho$ can be clearly seen in Fig.~\ref{fig5} (g), 
which is consistent with the previous results of Gaussian source~\cite{wtt2018,wtt2019,Zhoulong2020}. In Fig.~\ref{fig5} (a)-(i), with larger temperature or lower density, 
the difference of $C_{pp}(q)$ between different $T$ or $\rho$ becomes smaller, respectively. The Gaussian source radii are extracted for further discussion later in this article.

From the above plots, we can extract  $C_{max}(q)$, $i.e.$, the maximum value of $C_{pp}(q)$ as well as the full width at half maximum $(FWHM)$ of $C_{pp}(q)$ distribution,  $i.e.$ at 
$C_{pp}(q) = [C_{max}(q)-1]/2$. The source-temperature $T$ dependence of $C_{max}(q)$ and $FWHM$ for the proton-proton momentum correlation function with different density are given in Fig.~\ref{fig6}. As shown in Fig.~\ref{fig6} (a) and (b), both $C_{max}(q)$ and $FWHM$ decrease gradually with the increasing of $T$. In addition, both of them increase gradually with density. At high temperature, the change of $C_{max}(q)$ and $FWHM$ is very small and not plotted in the figure. Of course, the behavior of the $C_{max}(q)$ and $FWHM$ with $T$ and $\rho$ can also be clearly seen in Fig.~\ref{fig2}, and the increasing of  $C_{max}(q)$ and $FWHM$ are generally inversely proportional to Gaussian radius $r_{0}$ as shown later. 
Similarly, the system-size $A$ dependence of $C_{max}(q)$ and $FWHM$ for the proton-proton momentum correlation function at $T = 2 MeV$ and $\rho = 0.6\rho_{0}$ is shown in Fig.~\ref{fig7}. The dependence of $C_{max}(q)$ and $FWHM$ on system-size $A$ is quite similar to the temperature dependence in Fig.~\ref{fig6}. The $C_{max}(q)$ and $FWHM$ values become  smaller for larger systems.

Fig.~\ref{fig8} shows the source-temperature, density, and system-size dependence of Gaussian radii extracted from two-particle momentum correlation functions, 
where panels (a) and (b) are results with the smaller source size and the larger source size, respectively. The radii are extracted by a Gaussian source assumption, 
$i.e.$, $S(r) \approx \exp [-r ^2 /(4r^{2}_{0})]$, where $r_{0}$ is the Gaussian source radius from the proton-proton momentum correlation functions.
The theoretical calculations for $C_{pp}(q)$ was performed by using the $Lednick\acute{y}$ and $Lyuboshitz$ analytical method. The best fitting radius is judged by finding the minimum of the reduced chi-square between the $ThIQMD$ calculations and the Gaussian source assumption.
Since the effect of the strong $FSI$ scales as $f_c\left(k^*\right)/r^*$ in Eq.~$\left(9\right)$, one may read the sensitivity of the correlation function to the temperature $T$, density $\rho$ and atomic number $A$ from their effects on the Gaussian radius $r_{0}$. One may observe a linear dependence on these parameters up to $T \approx 8$ $MeV$ and then a lost of sensitivity in a plateau region at  higher temperatures in Fig.~\ref{fig8}.
As the density decreases, the decreasing speed of the Gaussian radius of the small system is larger than that of the larger system. Fig.~\ref{fig9} shows the Gaussian radius of the different system-size varies with the temperature in panels (a)-(c) or density in panels (d)-(f).
The Gaussian source radius is consistent with the system-size, $i.e.$, at higher temperature or larger density, the differences of Gaussian source between different system sizes are  bigger in the low density and low temperature region, but the difference in opposite conditions almost disappear. In other words, the sensitivity of the source radii to the system size seem to be different in the different regions of temperatures and densities. For example, the  sensitivity is   better in the region of lower $T$ and higher $\rho$ (Fig.~\ref{fig9}(b) and (c)), or it is better in the higher $T$ region for the lower $\rho$ (Fig.~\ref{fig9}(a)), or  it is better in the higher $\rho$ region for the lower $T$ (Fig.~\ref{fig9}(d)).

From the above discussion, it is demonstrated that the strength of the two-particle momentum correlation function is affected by the source temperature, density and system size.
The two-particle momentum correlation function strength is larger for a single source with lower temperature, higher density or smaller mass number as shown in Fig.~\ref{fig1}-\ref{fig5}. Otherwise, the strength becomes smaller. To some extents, the strong correlation between two particles is mainly caused by the closed position of each other in phase space in both coordinate and momentum.
Varying only one in the three condition parameters (temperature, density and system size), lower temperature means smaller momentum space, higher density means smaller coordinate space and small system size also mean smaller coordinate space to keep fixed density compared with large system size.  The dependencies of the two-particle momentum correlation function strength on the source temperature, density and system size could be explained by the change of the phase space sizes. Two particles emitted from small phase space will have strong correlation and those from large phase space will have weak correlation.  For example, the increase of the $C_{pp}(q)$ strength with the increase 
of the density for a fixed system size could be explained by the decreasing of the coordinate space as shown in Fig.~\ref{fig1} (a). And the small $C_{pp}(q)$ strength at temperature  higher than $8$ $MeV$ could be caused by the large momentum space compared with lower temperatures as shown in Fig.~\ref{fig1} (d-f).
The decrease of the $C_{pp}(q)$ strength with the increase of the system size for a fixed density could also be explained by the increasing of the coordinate space as shown in Fig.~\ref{fig5} (g). 
Thus it is concluded that the phase space size for the emitted nucleons have strong effect on strength of the two-particle momentum correlation function, 
which can also be seen in the extracted Gaussian radii as shown in Fig.~\ref{fig8}.  

\section{SUMMARY}

In summary, the two-particle momentum correlation functions for single excited sources are investigated using the Lednick$\acute{y}$ and Lyuboshitz analytical formalism 
with the phase-space information at the freeze-out stage for different initial temperatures and densities in a framework of the $ThIQMD$ transport approach. 
We mainly performed a series of studies focusing on the varied effects of source temperature, density and system-size on the two-particle momentum correlation functions.
The results reflect that the shape of the two-proton momentum correlation function is in accordance with the previous experimental data in heavy-ion collisions~\cite{Ghetti2000}. 
At the same time, the trend of the relationship between the two-proton momentum correlation and system-size is consistent with previous simulations~\cite{wtt2018,wtt2019,Zhoulong2020}. 
At low source-temperature, the larger density makes the two-particle momentum correlation stronger. However, at higher source temperature, the effect becomes 
almost disappear. Both proton-proton correlations and neutron-neutron correlations have the similar responses to temperature and density. This work also shows that the emission source is not much influenced by density above a certain temperature  
for a single excited source. In the same way, the emission source are softly influenced by temperature below a given density for a single excited source. 
In one word, the dependence of the two-particle momentum correlation function on the source temperature, density and system size could be explained by the change of the coordinate and/or momentum phase space sizes.
In the end, the Gaussian radii are extracted to explore the emission source sizes in single excited systems. Gaussian radii become larger in the larger systems. 
The dependence of the extracted Gaussian radius on source-temperature and density is consistent with behavior of the two-proton momentum correlation function as discussed in the texts.

\section*{Acknowledgments}
This work was supported in part by the National Natural Science Foundation of China under contract Nos.  11890710, 11890714, 11875066, 11925502, 11961141003, 11935001, 12147101 and 12047514,  the Strategic Priority Research Program of CAS under Grant No. XDB34000000, National Key R\&D Program of China under Grant No. 2016YFE0100900 and 2018YFE0104600,  Guangdong Major Project of Basic and Applied Basic Research No. 2020B0301030008, and the China PostDoctoral Science Foundation under Grant No. 2020M681140.

\end{CJK*}
\end{document}